\documentstyle[12pt,epsf,here]{article}
\newcommand{\PostScript}[7]{
\begin{figure}[H]
\begin{center}
\leavevmode
\epsfysize=#1cm
\vspace{#2cm}
\epsfbox{#3}
\par
\parbox{#5cm}{
\vspace{#4cm}
\caption[figure]{\renewcommand{\baselinestretch}{1} \small \normalsize #6}
\label{#7}}
\end{center}
\end{figure}
}

\begin{document}

\hfill{NORDITA-1999/36 CM}

\bigskip

\centerline{\large\bf Broken Symmetries in the Reconstruction of }
\centerline{\large\bf $\nu =1$ Quantum Hall Edges}

\medskip

\centerline{S.M. Reimann, M. Koskinen, S. Viefers$^{(*)}$, 
M. Manninen and B. Mottelson$^{(*)}$}

\medskip

\centerline{University of Jyv\"askyl\"a, PO Box 35,
FIN-40351 Jyv\"askyl\"a}

\centerline{$^{(*)}$ NORDITA, Blegdamsvej 17, DK-2100 Copenhagen}

\bigskip

\medskip

\centerline{\bf Abstract}

\begin{quote}
Spin-polarized reconstruction of the $\nu =1$ quantum Hall edge 
is accompanied by a spatial modulation of the charge density 
along the edge. We find that this is also the case for {\it finite} 
quantum Hall droplets: current spin density functional calculations 
show that the so-called Chamon-Wen edge forms a ring 
of apparently localized electrons around the maximum density 
droplet (MDD).
The boundaries of these different phases qualitatively 
agree with recent experiments.
For very soft confinement, Chern-Simons Ginzburg-Landau theory 
indicates formation of a non-translational invariant edge
with vortices (holes) trapped in the edge region.

PACS~{73.20.Dx, 73.61.-r, 85.30.Vw, 73.40.Hm}

\end{quote}

\noindent{\bf Introduction}

Edge states in the quantum Hall regime have been
subject to extensive study in recent years. In
particular, much interest has focused on
how the edge may reconstruct as the confining potential 
strength is varied (see [1] and Refs. therein). 
Various theoretical approaches, including Hartree-Fock methods, 
density functional theory, composite fermion
models and effective (mean field) theories have been used
to examine both small electron droplets (quantum dots) and
large quantum Hall systems, with and without spin.
In particular, many authors have been interested in edge reconstruction
of ferromagnetic quantum Hall states, including $\nu=1$ and
simple fractional (Laughlin) fillings.
Softening of the confining edge potential 
allows charge to move outward, and the edge may reconstruct.
How this happens, and whether or not the reconstruction
involves spin textures, depends on the relative strength 
of the electron-electron
interactions and the Zeeman energy, and on the steepness of 
the confining potential.
Much work has been based on Hartee-Fock techniques.
In 1994, Chamon and Wen found that the sharp $\nu=1$ edge of large 
systems or quantum dots may undergo a
polarized reconstruction to a ``stripe phase'' [2],
in which a lump of electrons becomes separated at a distance $\sim 2 l_B$
away from the original edge ($l_B= \sqrt{\hbar / eB}$). 
This reconstructed state is
translation invariant along the edge.
Using an effective sigma model and Hartree-Fock techniques
Karlhede {\em et al.} [3] then showed that Chamon and Wen's 
polarized reconstruction may be preempted
by edge spin textures if the Zeeman gap is sufficiently small.
Similar results were obtained by Oaknin {\it et al.}~[4]
for finite quantum Hall droplets.
Spin textures are configurations of the spin field
in which the spins tilt away from their bulk direction
on going across the edge; on going {\em along} the edge,
they precess about the direction of the external field
with some wave vector $k$.
The edge textures posess a topological density which
can be shown to be proportional to the electron density.
Thus, one may say that tilting spins moves charge,
which is why edge spin textures represent a mechanism
for edge reconstruction.
Later it turned out~[1,5], that the Chamon-Wen edge is, in fact, 
unstable: a polarized reconstruction with a {\em modulated charge density} 
along the edge is always
lower in energy than the translation invariant Chamon-Wen
edge. 
Numerical (Hartree-Fock) studies of the ground state,
together with an analysis of the softening of low-energy 
edge modes at weak confining potentials,
have resulted in a phase diagram [1,5], giving the following
picture of the $\nu=1$ edge:
For very steep confining potentials, the edge is sharp
and fully polarized. Upon softening of the confining potential,
the edge will {\em either} reconstruct into a spin textured state
with a translation invariant charge density along the edge
(for small Zeeman gaps) {\em or} into a polarized charge density wave edge 
(for large Zeeman gaps). For even softer confining potentials and 
sufficiently small Zeeman gaps a combination of charge modulation 
along the edge and spin textures may occur.

\bigskip

\noindent{\bf Broken-symmetry edge states in quantum dots} 

The above mentioned phases of edge reconstruction can also occur in 
{\it finite} quantum Hall systems such as quantum dots.
The spin-textured edge exists only for sufficiently smooth confinement 
and small enough Zeeman coupling. We restrict the following discussion 
to the spin-polarized regime.
In a strong enough magnetic field, the electrons 
fill the lowest Landau level: the so-called 
maximum density droplet (MDD) [6] is formed, in which the electrons 
occupy adjacent orbitals with 
consecutive angular momentum. The MDD is the finite-size analogue to the 
bulk $\nu =1$ quantum Hall state with an approximately constant density at
its maximum value $(2\pi l_B^2)^{-1}$.
Increasing the magnetic field effectively compresses the electron droplet.
At a certain field strength, the dense 
arrangement of electrons costs too much Coulomb energy. The droplet 
then takes advantage of moving electrons from lower to higher angular 
momentum states and re-distributes its density~[2]. 
This, however, may occur together with a breaking of the rotational symmetry 
in the internal coordinates of the many-body wave function.
The self-consistent mean-field 
solution can show such {\it intrinsic} symmetry breaking. 
The latter implies the occurence of a rotational 
band which can be obtained by projection.
For filling factors around $\nu =1$ we 
apply current spin density functional theory (CSDFT) [8] to calculate the 
ground-state densities of $N$ parabolically confined electrons, 
avoiding any spatial symmetry restrictions of the solutions. 
For the technical details of the calculations, 
we refer to [7].
An example for the
edge reconstruction in finite quantum Hall droplets is shown in 
the left of Fig.~1 (see next page) for $N=42$ electrons. In the ground state 
the MDD is stable up to a field of about $2.6$~T.
At about $2.7$~T, reconstruction has taken place: at a distance of  
$\sim 2 l_B$ from the remaining inner (smaller) MDD, 
a ring of separate lumps of charge density is formed, with each lump 
containing one electron and having a radius somewhat larger than the 
magnetic length $l_B$. 
Goldmann and Renn [9] recently suggested crystallized edge states 
which appear similar to the reconstructed edges within CSDFT.
For still higher fields, the sequential 
formation of ring-like edges continues until the whole droplet is fully 
reconstructed [7,10]. The apparent localization at the edge
is accompanied by a narrowing of the 
corresponding band of single-particle energy levels. The existence of the   
inner MDD surrounded by the broken-symmetry edge opens up the possibility
to observe rotational spectra of the edge.
We next study the formation of the MDD and its reconstruction systematically,
varying both particle number $N$ and magnetic field $B$. For fixed $N$
we keep the average electron density constant. 
Changing the field $B$ has a similar effect on the reconstruction
as varying the softness of the external confinement:
a higher field compresses the droplet.
At constant strength of the oscillator for fixed particle number, 
but larger field the confinement then is effectively weaker. 
We obtain a phase diagram as a function of the number $N$ 
of confined electrons and the field $B$, which is schematically 
shown in Fig.~1. (For more details see [7]).
With increasing $N$, the polarization line which separates 
the fully polarized MDD states from the unpolarized states 
approaches the reconstruction line. The latter separates the MDD regime 
from the Chamon-Wen (CW) edge formation. This is schematically indicated 
by the dashed lines in Fig.~1.
Note that the shapes of these phase boundaries differ from the results of  
Ferconi and Vignale [11], as they used a fixed confinement 
strength for different dot sizes.
In recent experiments [12] a phase diagram was obtained 
from addition energy spectra measured as a function of magnetic field.
The phase boundaries qualitatively agree with the results obtained 
from the CSDFT calculations, if the average electron density is 
kept constant. Its value determines the magnetic field strength at 
which the phase transitions occur: Increasing the density shifts 
the phase boundaries to higher $B$-values.

\vskip-1.5truecm
\noindent
\PostScript{7}{0}{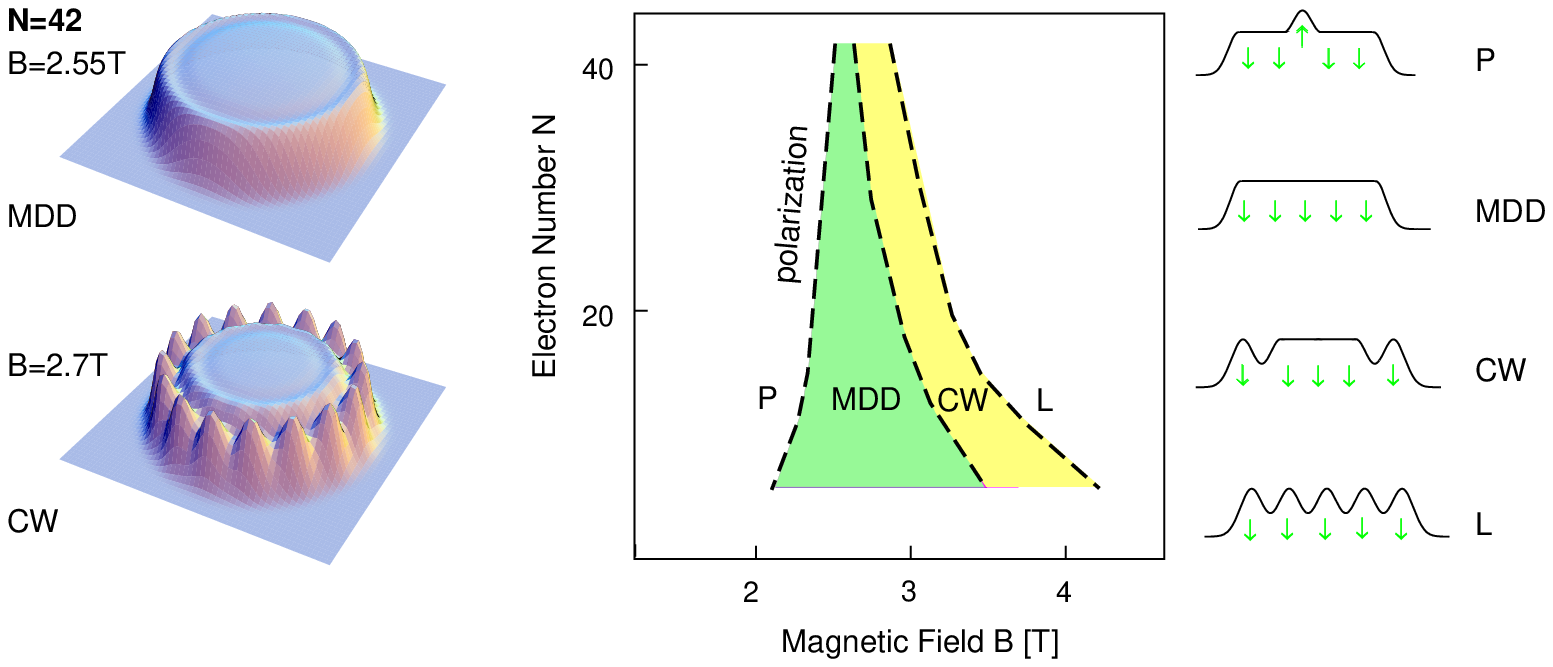}{0}{15}{\normalsize 
{\it Left:} Charge density of a spin-polarized MDD before ($B=2.55$~T)
and after ($B=2.7$~T) edge reconstruction.
{\it Right:} Phase diagram of parabolic quantum dots 
at a density roughly corresponding to a value
$1/(\pi r_s^2)$, where $r_s=2a_B^*$ and 
$a_B^*$ is the effective Bohr radius.
The schematic density profiles on the right indicate the different phases of 
polarization (P), formation of the maximum density droplet (MDD),
edge reconstruction (CW) and localization (L) }{fig1}
\vskip-0.5cm
\noindent{\bf Edge reconstruction within CSGL theory}

Turning to filling fractions $\nu \le 1$, we now study the {\it infinite}, 
straight
quantum Hall edge within the framework of Chern-Simons Ginzburg-Landau
(CSGL) theory [13]. This is an effective (mean field) model of the
FQHE, based on the concept of ``statistical transmutation'': It
models the electrons at $\nu=1/(2m+1)$ as {\em bosons}, each carrying 
$2m+1$ quanta of (``statistical'') flux; in the mean field sense, this
statistical field is cancelled by the external magnetic field,
making the $\nu=1/(2m+1)$ quantum Hall state equivalent to a system of
charged bosons in {\em zero} magnetic field. 
This model has proven quite 
successful in describing bulk properties of the FQHE.
The edge can be studied by solving the CSGL field equations in
the presence of an external confining potential. 
Leinaas and Viefers~[14] recently showed the existence of edge spin textures
in this model for soft enough confining potentials and Zeeman energies
smaller than some critical value,
in qualitative agreement with previous work~[3,4].
Fig.~2 shows such a solution in the limit where the minority spin
density is small. 
As mentioned, the charge density of the spin textured edge is translation
invariant along the edge. The CSGL studies [14] further
indicate the possibility of another kind of edge reconstruction, at 
even softer confining
potentials, to a non-translation invariant edge with vortices (holes) 
trapped in the edge region. Several authors have adressed this type
of reconstruction (see [5] and references therein).

\vskip-1truecm
\noindent
\PostScript{4}{0}{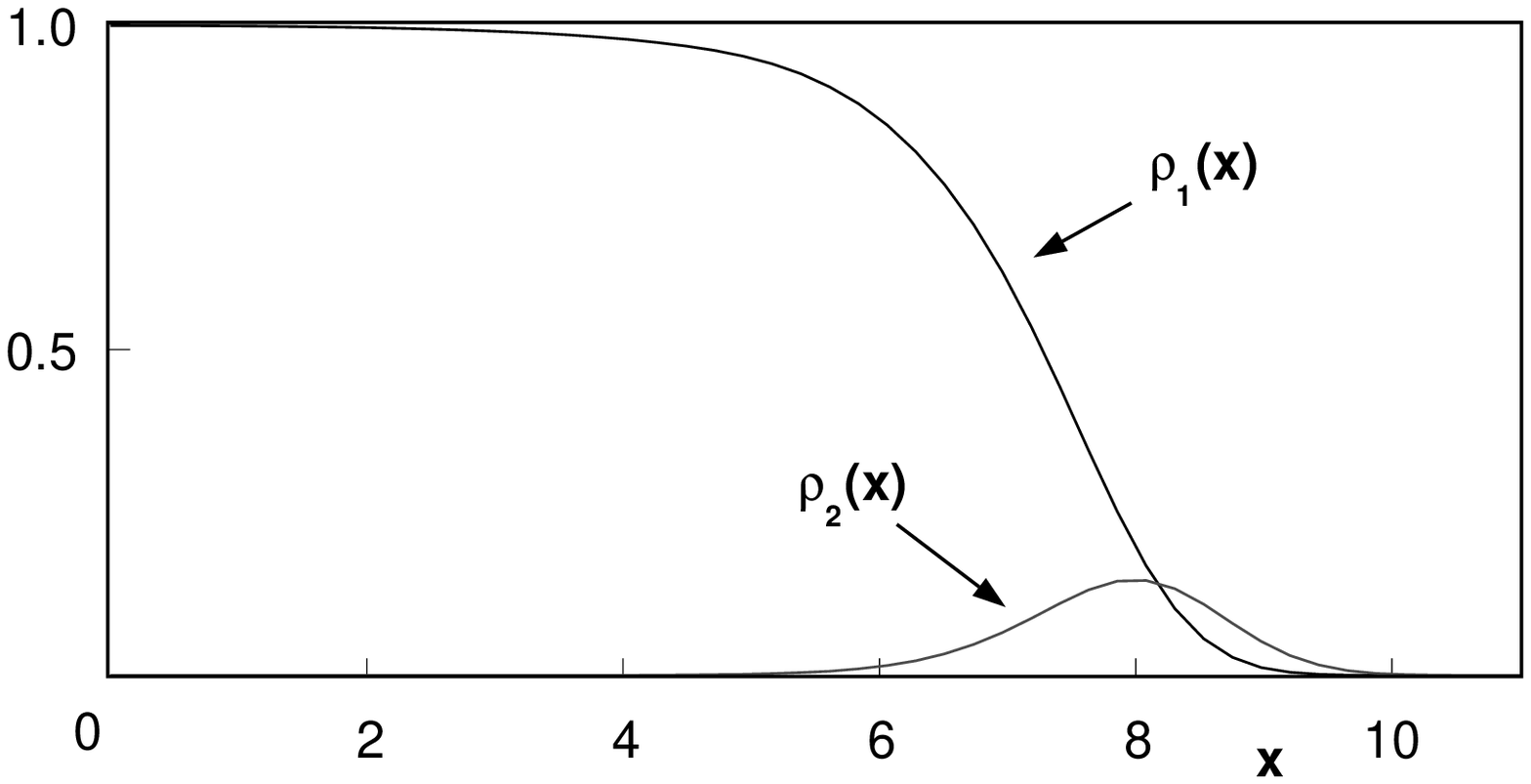}{0}{15}{
Spin textured edge in CSGL theory:
Majority- ($\rho_1$) and minority ($\rho_2$) spin density
profiles close to the critical Zeeman gap for a harmonic
confining potential of the form
$V(x) = 0.05~\theta(x-5)\cdot (x-5)^2$ 
(with $x$ in units of the magnetic length $l_B$). 
The spin unit vector $\hat n$ is given by $n_x+in_y=\sqrt {
1-n_z(x)^2}~e^{iky}$ and $n_z(x) = (\rho_1(x)-\rho_2(x))/(\rho_1(x)+\rho_2(x))$.
The critical
Zeeman gap is approximately $0.06~ \hbar\omega_c$,
and the wave vector of spin precession along the edge is
$k = 1.0 ~l_B^{-1}$. $\rho_1$ is normalized by its constant
bulk value}{fig2}

\vskip-0.5truecm
\noindent
{\bf References}

\smallskip

\noindent
[1]~A. Karlhede and K. Lejnell, Physica E {\bf 1}, 41 (1997).

\noindent
[2]~C. de C. Chamon and X.G. Wen, 
                  {Phys. Rev.} B{\bf 49}, 8227 (1994).

\noindent
[3]~A. Karlhede {\em et al.},
                  {Phys. Rev. Lett.} {\bf 77}, 2061 (1996).

\noindent
[4]~J.H. Oaknin {\it et al.}, Phys. Rev. B {\bf 54}, 16850 (1996);
{\bf 57}, 6618 (1998).

\noindent
[5]~M. Franco and L. Brey,
                  {Phys. Rev.} B{\bf 56} 10383 (1997).

\noindent
[6] A. H. McDonald, S.R.E. Yang, M. D. Johnson, Aust. J. Phys. {\bf 46}, 345 (1993).

\noindent
[7] S.M. Reimann, M. Koskinen, M. Manninen and B. Mottelson, cond-mat/9904067. 

\noindent
[8] G. Vignale and M. Rasolt, Phys. Rev. B{\bf 37}, 10685 (1988).

\noindent
[9] E. Goldmann and S. Renn, to be published.

\noindent
[10] H.-M. M\"uller and S.E. Koonin, Phys. Rev. B{\bf 54}, 14532 (1996).

\noindent
[11] M. Ferconi and G. Vignale,  Phys. Rev. B{\bf 56}, 12108 (1997).

\noindent
[12] T. H. Oosterkamp {\it et al.}, Phys. Rev. Lett. {\bf 82}, 2931 (1999).

\noindent
[13] S.C. Zhang {\em et al.},
               {Phys. Rev. Lett.} {\bf 62}, 82 (1988);
               D.H. Lee and C.L. Kane,
               {Phys. Rev. Lett.} {\bf 64}, 1313 (1990). 

\noindent
[14] J.M. Leinaas and S. Viefers, 
                  {Nucl. Phys. B}{\bf 520}, 675 (1998).

\end{document}